\documentclass[aps,prf,reprint,onecolumn,superscriptaddress, 12pt]{revtex4}

\usepackage{graphicx}
\usepackage{xcolor}
\usepackage{units}

\usepackage{amsmath}

\begin{document}

\title{The asymmetric coalescence of two droplets with different surface tensions is caused by capillary waves}

\author{Michiel A. Hack}
\email{m.a.hack@utwente.nl}
\affiliation{Physics of Fluids Group, Max Planck Center Twente for Complex Fluid Dynamics, Faculty of Science and Technology, University of Twente, P.O. Box 217, 7500 AE Enschede, The Netherlands}

\author{Patrick Vondeling}
\affiliation{Physics of Fluids Group, Max Planck Center Twente for Complex Fluid Dynamics, Faculty of Science and Technology, University of Twente, P.O. Box 217, 7500 AE Enschede, The Netherlands}

\author{Menno Cornelissen}
\affiliation{Physics of Fluids Group, Max Planck Center Twente for Complex Fluid Dynamics, Faculty of Science and Technology, University of Twente, P.O. Box 217, 7500 AE Enschede, The Netherlands}

\author{Detlef Lohse}
\email{d.lohse@utwente.nl}
\affiliation{Physics of Fluids Group, Max Planck Center Twente for Complex Fluid Dynamics, Faculty of Science and Technology, University of Twente, P.O. Box 217, 7500 AE Enschede, The Netherlands}
\affiliation{Max Planck Institute for Dynamics and Self-Organization, 37077 G\"ottingen, Germany}

\author{Jacco H. Snoeijer}
\email{j.h.snoeijer@utwente.nl}
\affiliation{Physics of Fluids Group, Max Planck Center Twente for Complex Fluid Dynamics, Faculty of Science and Technology, University of Twente, P.O. Box 217, 7500 AE Enschede, The Netherlands}

\author{Christian Diddens}
\email{c.diddens@utwente.nl}
\affiliation{Physics of Fluids Group, Max Planck Center Twente for Complex Fluid Dynamics, Faculty of Science and Technology, University of Twente, P.O. Box 217, 7500 AE Enschede, The Netherlands}
\affiliation{Department of Mechanical Engineering, Eindhoven University of Technology, P.O. Box 513, 5600 MB Eindhoven, The Netherlands}

\author{Tim Segers}
\email{t.j.segers@utwente.nl}
\affiliation{BIOS Lab on a Chip Group, Max Planck Center Twente for Complex Fluid Dynamics, MESA+ Institute for Nanotechnology, University of Twente, Enschede, The Netherlands}
\affiliation{Physics of Fluids Group, Max Planck Center Twente for Complex Fluid Dynamics, Faculty of Science and Technology, University of Twente, P.O. Box 217, 7500 AE Enschede, The Netherlands}

\date{\today}

\begin{abstract}
When two droplets with different surface tensions collide, the shape evolution of the merging droplets is asymmetric. Using experimental and numerical techniques, we reveal that this asymmetry is caused by asymmetric capillary waves, which are the result of the different surface tensions of the droplets. We show that the asymmetry is enhanced by increasing the surface tension difference, and suppressed by increasing the inertia of the colliding droplets. Furthermore, we study capillary waves in the limit of no inertia. We reveal that the asymmetry is not directly caused by Marangoni forces. In fact, somehow counterintuitive,  asymmetry is strongly reduced by the Marangoni effect. Rather, the different intrinsic capillary wave amplitudes and velocities associated with the different surface tensions of the droplets lie at the origin of the asymmetry during droplet coalescence. 
\end{abstract}

\maketitle

\section{Introduction}
The collision and subsequent coalescence of liquid droplets is omnipresent in both nature and technology. For example, small droplets collide in the atmosphere, forming larger ones that eventually fall to the Earth as rain \cite{Grabowski2013}. The coalescence of ink droplets is also a vital process in inkjet printing \cite{Lohse2022, Wijshoff2010, Wijshoff2018}, the collisions and coalescence of ethanol and diesel droplets play an important role in combustion engines \cite{Qian1997, Wang2005, Chen2007}, and the in-air coalescence of droplets has recently been introduced as a method to mass production of functional micromaterials \cite{Visser2018, Jiang2021}. The fundamental importance of droplet collision and coalescence has led to intense investigation in recent decades \cite{Eggers1999, Duchemin2003, Roisman2004, Ristenpart2006, Paulsen2014}. These studies found that the collision and coalescence dynamics of identical droplets are determined by both the geometry and composition of the droplets  \cite{Paulsen2011, HernandezSanchez2012, Eddi2013}. It is therefore expected that these dynamics are strongly altered when the two droplets are not identical. Indeed, unequal-sized droplets show different collision outcomes when compared to collisions between droplets of equal size \cite{Zhang2009, Tang2012}. Moreover, when two droplets of different surface tensions coalesce, they show fundamentally different dynamics than their equal-surface tensions counterpart---they exhibit intricate phenomena such as `delayed coalescence' \cite{Karpitschka2014, Karpitschka2012, Borcia2011, Bruning2018, Karpitschka2010} and have enhanced internal mixing \cite{Blanchette2010}.

Here, we study the mid-air collision of two droplets with different surface tensions. The difference between the surface tensions of the droplets induces a tangential Marangoni stress at the droplet interface, which results in the engulfment of the higher surface tension droplet by the lower surface tension liquid \cite{Planchette2010, Koldeweij2019}. A similar system was studied by Gao \emph{et al.} \cite{Gao2005}, and later by Kohno \emph{et al.} \cite{Kohno2013, Suzuki2014}, who reported that the shape of such droplets during their coalescence is asymmetric. Gao \emph{et al.} and Kohno \emph{et al.} attribute the asymmetric droplet shape to capillary wave interference \cite{Gao2005, Kohno2013}. In their model, a capillary wave forms at the point of contact of the two droplets, and subsequently travels over the high surface tension droplet's interface, driven by the difference in surface tensions between the two droplets, i.e., due to the Marangoni effect. When the capillary wave reaches the droplet apex, it constructively interferes with itself, forming a protrusion, thus resulting in an asymmetric droplet shape. Capillary waves also form and interfere during the coalescence of identical droplets, though in that case the protrusions that appear are identical on both droplets \cite{Paulsen2014, Billingham2005, Duchemin2003, Thoroddsen2007, Zhang2009}. Understanding the dynamics of engulfment and capillary waves in the presence of the Marangoni effect, and in particular their interaction in droplet collisions, remains an open question.

This work offers a detailed study of the asymmetric evolution of the droplet shapes during the collision of two droplets with different surface tensions. Using a combination of experiments and numerical simulations, we show, presumably unsurprisingly, that the asymmetry grows as the difference between the surface tensions of the two droplets is increased. However, more remarkably, we show that this is not due to the Marangoni forces. Instead, we unravel the critical role of capillary waves in the formation of the asymmetric droplet shapes. Finally, we also reveal that the asymmetry can be suppressed by increasing the collisional Weber number. 

\begin{figure}[b!]
	\begin{center}
		\includegraphics[scale=1.1]{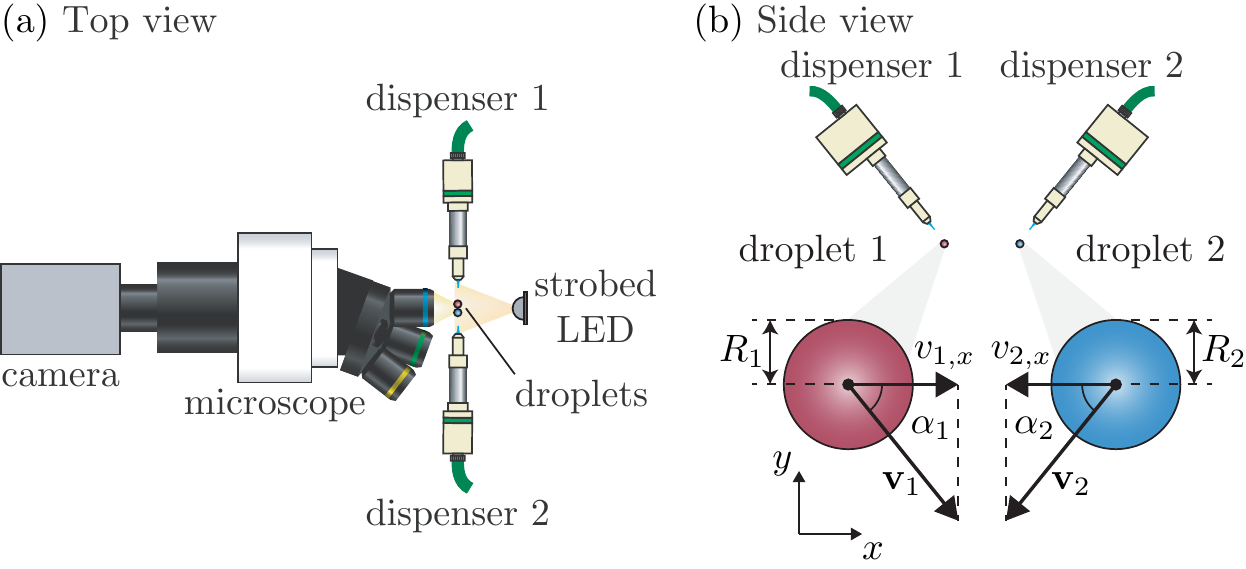}
		\caption{(a) Schematic of the experimental setup  (top view). (b) Definitions of the collision parameters (side view).}
		\label{collisions_fig1}
	\end{center}
\end{figure}

\section{Methods}
\label{sec:methods}
\subsection{Experimental method}
The experimental setup is shown in Fig.~\ref{collisions_fig1}a. Two piezo-electric droplet dispensers (AD-K-501, Microdrop Technologies) were used to generate droplets that collide in flight. A continuous stream of droplets was generated to prevent compositional changes in the liquid due to evaporation and fouling of the liquid interface at the nozzle exit of the dispenser. A high-speed camera (HPV-X2, Shimadzu) equipped with a 20$\times$ microscope lens (Olympus) was used to record the dynamics of the collisions with a spatial resolution of 1.6~$\mu$m/pixel at one to two millions frames per second. We used a strobed LED to generate high-intensity light flashes with a duration of $300$~$\mu$s to illuminate the collisions. The duration of a single LED flash is longer than the collision time. The droplet dispensers were oriented at an angle to facilitate collisions in the imaging plane of the microscope. The impact velocity (see Fig.~\ref{collisions_fig1}b) is therefore defined as $v = v_{1,x} + v_{2,x} = \cos \alpha_1 \lVert {\bf v}_1 \rVert + \cos \alpha_2 \lVert {\bf v}_2 \rVert$, where $\alpha_1 \approx \alpha_2$ and $v_{1,x} \approx  v_{2,x}$. The velocities $\lVert {\bf v}_1 \rVert$ and $\lVert {\bf v}_2 \rVert$ as well as the angles $\alpha_1$ and $\alpha_2$ were extracted from the movie frames captured prior to the collision event. The velocity of the droplets was controlled by the amplitude of the electric pulse applied to the piezo-electric actuator in the dispenser, resulting in impact velocities in the range 0.4~m/s~$\leq$~$v$~$\leq$~6.7~m/s. The radiii of the droplets were $R_1~\approx~R_2~\approx~35$~$\mu$m in all experiments. 

The experiments were performed with water (MilliQ, Millipore Corporation), ethanol (99.8\% purity, Sigma Aldrich), and mixtures of water and ethanol. The properties of these liquids are summarized in Table~\ref{tab:liquidProperties}. Water and ethanol were chosen for their large difference in surface tension, low viscosities, and miscibility. An estimated 0.8\% of the initial droplet volume evaporates during the flight and collision of the droplets, such that the surface tensions of the droplets remain approximately constant during the experiment, see the Supplementary Material for details.

\begin{table}[t!]
\begin{center}
\caption{\label{tab:liquidProperties}%
Properties of the liquids used in the experiments of this study. The surface tensions are based on the values reported in Ref.~\cite{Vazquez1995}. The density and viscosity are based on the values reported in Ref.~\cite{Gonzalez2007}.}
\begin{tabular}{ l @{\qquad} c @{\qquad} c @{\qquad} c}
\hline
\textrm{Liquid} & $\gamma$ [mN/m] & $\eta$ [mPa$\cdot$s] & $\rho$ [kg/m$^3$]
\\
\hline
water 					& 	72.5		& 	0.92		& 	997	\\ 
8.0 wt\% ethanol in water 		& 	51.3		&	1.51		&	986	\\
34.5 wt\% ethanol in water	& 	31.9		& 	2.56		& 	941	\\
ethanol 					& 	22.9 		& 	1.17		& 	789	\\
\hline
\end{tabular}
\end{center}
\end{table} 

\subsection{Dimensionless groups}
We define the dimensionless surface tension difference $\tilde{\gamma}$ and collisional Weber number as
\begin{equation}
\tilde{\gamma}~=~\frac{\gamma_2-\gamma_1}{\gamma_1}, \quad \text{and} \quad \mathrm{We}~=~\frac{\bar{\rho}v^2\bar{R}}{\gamma_2},
\end{equation} 
respectively, where $\bar{\rho}~=~(\rho_1+\rho_2)/2$, $\rho_1$ and $\rho_2$ are the densities, and $\gamma_1$ and $\gamma_2$ the surface tensions of droplets 1 and 2, respectively, and $\bar{R}$ is the average radius of the droplets. We always choose $\gamma_2~>\gamma_1$ by definition. In the present work, $\tilde{\gamma}$ and $\mathrm{We}$ were varied over the range of $0~\leq~\tilde{\gamma}~\leq~4$, and $0~\leq~\mathrm{We}~\leq~25$. The density ratio $\rho_2/\rho_1$ is an additional dimensionless parameter, which is determined by the values in Table~\ref{tab:liquidProperties}.

The Ohnesorge number and Bond number are defined as 
\begin{equation}
\mathrm{Oh}~=~\frac{\eta_2}{\sqrt{\bar{\rho} \gamma_2 \bar{R}}}, \quad \text{and} \quad \mathrm{Bo}~=~\frac{(\bar{\rho}-\rho_\mathrm{air})g\bar{R}}{\gamma_2},
\end{equation} 
respectively, where $\eta_2$ is the viscosity of droplet 2, $g$ is the gravitational acceleration, and $\rho_\mathrm{air}$ is the density of air. The gravitational and viscous forces can be neglected in all experiments and numerical simulations considered in the present work, since $\mathrm{Oh}~=~\mathcal{O}(10^{-2})~\ll~1$ and $\mathrm{Bo}~=~\mathcal{O}(10^{-4})\ll~1$. 

Finally, time is normalized by the inertio-capillary time scale
\begin{equation}
\tilde{t}~=~\frac{t}{\tau}, \quad \text{with} \quad \tau_i~=~\sqrt{\frac{\rho_i R_i^3}{\gamma_i}},
\end{equation}
which is associated with the surface tension-driven deformation of the droplets during coalescence \cite{Thoroddsen2000}. Here, $i~\in~[1,2]$ indicates that the properties of either droplet 1 or 2 are used, based on which is being considered. 

\subsection{Numerical method and its validation}
\label{sec:collisions_numericalmethod}
A numerical simulation of the collision process demands an accurate tracking of the interface and the treatment of the composition and velocity field, which are strongly coupled due to the Marangoni effect.
To that end, an axisymmetric sharp-interface arbitrary Lagrangian-Eulerian (ALE) finite element method is employed. Both droplets are represented by a mesh consisting of triangular elements, using second order basis functions for the velocity $\mathbf{u}$ and linear basis functions for the pressure $p$ and composition $c$. Due to the low viscosity and density ratio, the gas phase is not considered. At higher impact velocities, the presence of the gas phase can however influence the collision due to the build-up of an air cushion, the formation of a dimple, and the presence of entrapped bubbles \cite{Thoroddsen2003, Tran2013, Hendrix2016}. At lower collision velocities, however, hardly any dimple forms in collisions, and simulations with and without consideration of the gas phase lead to almost identical results.
Evaporation is also disregarded in the numerical simulations, given the short time scale of the collision process compared to typical droplet evaporation time scales.

The Navier-Stokes equations and the convection-diffusion equation for the composition are solved, i.e.,
\begin{align}
\rho\left(\partial_t\mathbf{u}+\mathbf{u}\cdot\nabla\mathbf{u}\right)&=-\nabla p +\nabla\cdot\left[\eta\left(\nabla\mathbf{u}+(\nabla\mathbf{u})^\mathrm{T}\right)\right],\\
\partial_t \rho + \nabla\cdot\left(\rho\mathbf{u}\right)&=0,\\
\rho\left(\partial_t c + \mathbf{u}\cdot\nabla c\right)&=\nabla\cdot\left(\rho D\nabla c\right)\,. 
\end{align}
where $D$ is the diffusion coefficient. The properties $\rho$, $\eta$, $D$ and also the surface tension $\gamma$ are functions of the composition $c$.
At the free interface, the Laplace pressure and the Marangoni shear stress are imposed, i.e.,
\begin{align}
\mathbf{n}\cdot\mathbf{T}\cdot \mathbf{n}=\kappa\gamma, \quad \text{and} \quad \mathbf{n}\cdot\mathbf{T}\cdot\mathbf{t}=\nabla_\mathrm{S}\gamma\cdot\mathbf{t},
\end{align} 
with the stress tensor $\mathbf{T}=-p\mathbf{1}+\eta(\nabla\mathbf{u}+(\nabla\mathbf{u})^\mathrm{T})$, the normal $\mathbf{n}$, and tangent $\mathbf{t}$. Here, $\kappa$ is the curvature of the interface and $\nabla_\text{S}$ is the surface gradient operator.
Finally, the kinematic boundary condition is enforced via a field of Lagrange multipliers, which ensures that the normal mesh motion at the interface coincides with the normal velocity $\mathbf{u}\cdot\mathbf{n}$, whereas the positions of the mesh nodes in the bulk follow the motion of a deformed pseudo-elastic body \cite{Cairncross2000, Heil2006}.

The coalescence is initiated as follows. The moment the droplets approach each other within $1\%$ of the average radius, the interfaces of both droplets are connected and the mesh is reconstructed, followed by an interpolation of the solved fields and their values of the previous time step, which is required for the second order time stepping via the backward differentiation formula. Note that $R_1~=~R_2$ unless otherwise stated. 
Mesh reconstruction is also invoked whenever the mesh quality suffers from strong deformations.

\begin{figure}[t!]
	\begin{center}
		\includegraphics[scale=1.1]{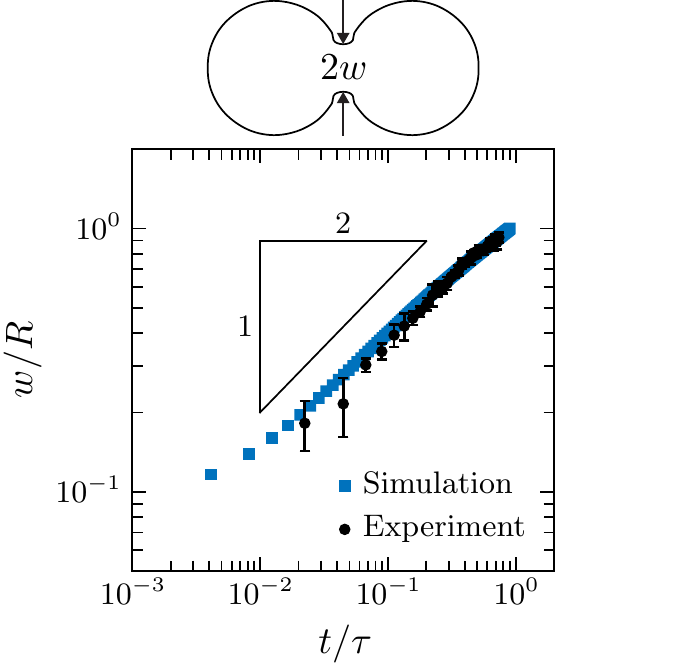}
		\caption{Temporal evolution of the neck width $w$ for two coalescing water droplets ($\gamma~=~72.5$~mN/m, $\rho~=~997$~kg/m$^3$, and $\eta~=~0.92$~mPa$\cdot$s). The experimental data is an average of five measurements with $\mathrm{We}~=~0.17$. The error bars indicate the standard deviation between the measurements at each time step.}
		\label{collisions_fig2}
	\end{center}
\end{figure}

The method is implemented on the basis of the framework \textsc{oomph-lib} \cite{Heil2006}. All equations are solved with a monolithic Newton method using a direct solver for the inversion of the Jacobian. The implementation has already been successfully applied for various scenarios involving multi-component droplet dynamics and Marangoni flow, see e.g., Refs. \cite{Li2019b, Li2019}.
The dependence of the results on the mesh size and the time step is discussed in the Supplementary Material~\ref{sec:appendix_timestepgridresolution}.

\begin{figure}[h!]
	\begin{center}
		\includegraphics[scale=1.1]{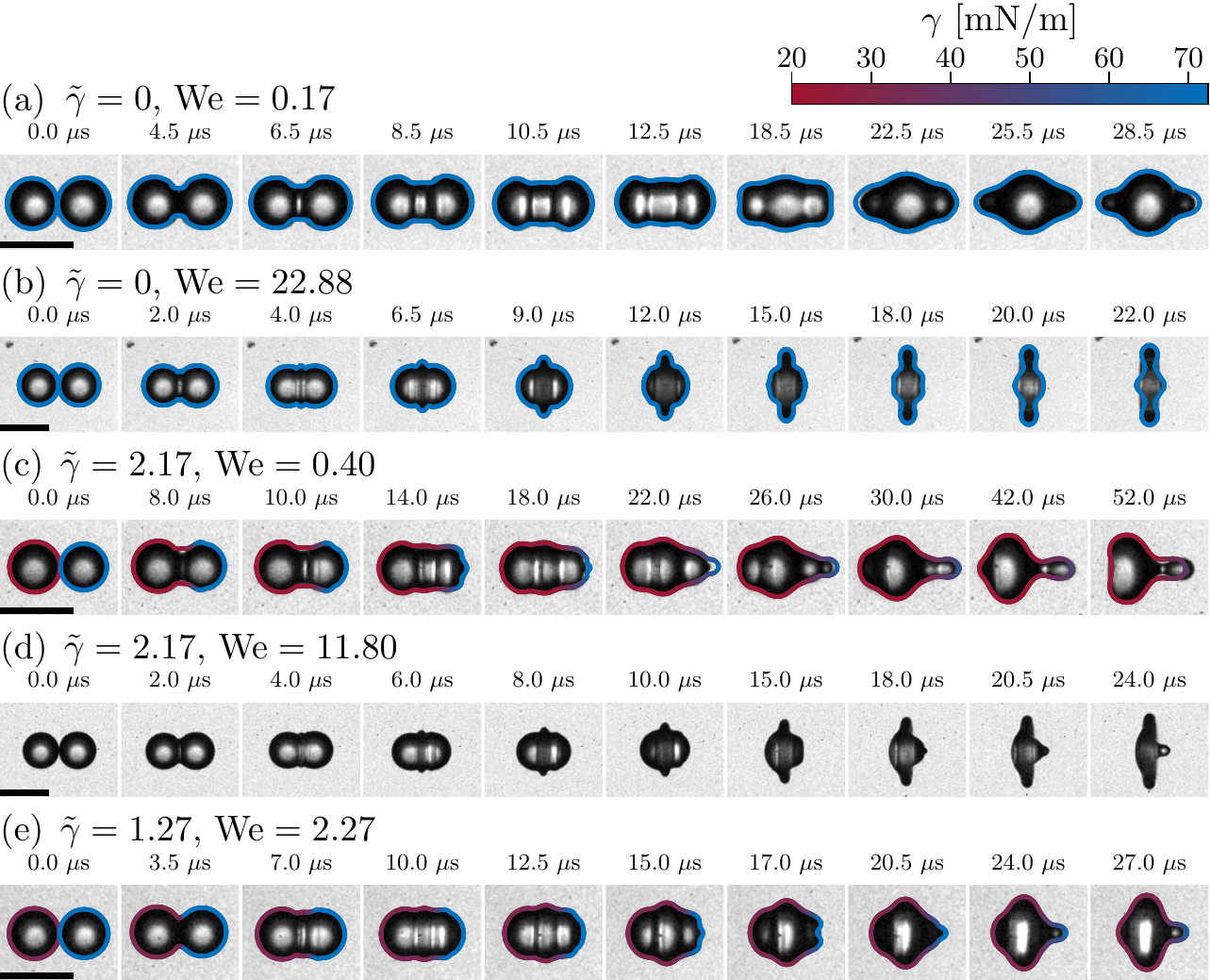}
		\caption{Direct comparison of experiments (images) and numerical simulations (colored lines). The experimental parameters $v_i$, $R_i$, $\gamma_i$, $\eta_i$, and $\rho_i$ ($i~\in~[1,2]$) were used as input for the numerical simulations. Collisions between (a) two water droplets with $\tilde{\gamma}~=~0$, $\mathrm{We}$~=~0.17, $\mathrm{Bo}$~=~1.6$\times$10$^{-4}$, $\mathrm{Oh}$~=~1.8$\times$10$^{-2}$, (b) two water droplets with $\tilde{\gamma}~=~0$, $\mathrm{We}$~=~22.88, $\mathrm{Bo}$~=~2.2$\times$10$^{-4}$, $\mathrm{Oh}$~=~1.7$\times$10$^{-2}$, (c) a water droplet and an ethanol droplet with $\tilde{\gamma}~=~2.17$, $\mathrm{We}$~=~0.40, $\mathrm{Bo}$~=~1.3$\times$10$^{-4}$, $\mathrm{Oh}$~=~2.0$\times$10$^{-2}$, (d) a water droplet and an ethanol droplet (no simulations available, see the discussion in section \ref{sec:collisions_numericalmethod}) with $\tilde{\gamma}~=~2.17$, $\mathrm{We}$~=~11.80, $\mathrm{Bo}$~=~1.7$\times$10$^{-4}$, $\mathrm{Oh}$~=~1.9$\times$10$^{-2}$, (e) a water droplet and a 34.5 wt\% ethanol droplet with $\tilde{\gamma}~=~1.27$, $\mathrm{We}$~=~2.27, $\mathrm{Bo}$~=~1.5$\times$10$^{-4}$, $\mathrm{Oh}$~=~1.9$\times$10$^{-2}$. See the Supplementary Material for the movies of these collisions.}
		\label{collisions_fig3}
	\end{center}
\end{figure}

Figures \ref{collisions_fig2} and \ref{collisions_fig3} show the validation of our numerical method against our experimental results. The temporal evolution of the neck width $w$ for $\mathrm{We}~=~0$ and $\tilde{\gamma}~=~0$ is shown in Fig.~\ref{collisions_fig2}, and it is in good agreement with the expected scaling law $w \propto (\gamma R/\rho)^{1/4} t^{1/2}$ \cite{Eggers1999}. For comparison, we show the experimentally obtained neck width for $\mathrm{We}~=~0.17$ (due to the nature of the setup, we are unable to achieve $\mathrm{We}~=~0$ in the experiments) and find good agreement with the numerical simulations. A comparison of the coalescence dynamics for various $\mathrm{We}~>~0$ and $\tilde{\gamma}$ is shown in Fig.~\ref{collisions_fig3}, where the experimental parameters $v_i$, $R_i$, $\gamma_i$, $\eta_i$, and $\rho_i$ ($i~\in~[1,2]$) were used as input for the numerical simulations. A detailed physical discussion of Fig~\ref{collisions_fig3} is provided in section~\ref{sec:asymmetricdropletcollisions}. The surface tension depends on the local composition $c$ as $\gamma = 0.016367+0.007399/(c+0.131824)$ (in N/m), which is an empirical fit to the data presented by Vazquez \emph{et al.} \cite{Vazquez1995}. The density $\rho~=~997.0+(-208.0(0.660c+0.340c^2))$ (in kg/m$^3$) \cite{Gonzalez2007}, viscosity $\eta~=~0.000923+0.00821c-0.0111c^2+0.00314c^3$ (in Pa$\cdot$s) \cite{Gonzalez2007}, and diffusion coefficient $D~=~1.25\times10^{-9}(1.0-2.78c+2.72c^2)$ (in m$^2$/s) \cite{Parez2013} were similarly obtained by empirical fits. Here, $c~=~0$ indicates pure water and $c~=~1$ indicates pure ethanol. Excellent agreement between the experiments and numerical simulations demonstrates that our numerical method accurately describes the collision dynamics. No simulation is shown in Fig.~\ref{collisions_fig3}d, because the formation of a toroidal bubble (which forms when the capillary waves coalesce with the opposing droplet) causes the simulation to fail \cite{Eggers1999}. 

\section{Collisions between droplets of different surface tensions} 
\label{sec:asymmetricdropletcollisions}
We first discuss our results related to the collision between two droplets with different surface tensions at finite $\mathrm{We}$. Figure~\ref{collisions_fig3} gives an overview of the collision dynamics observed for various combinations of $\tilde{\gamma}$ and $\mathrm{We}$. A collision between two identical droplets ($\tilde{\gamma}~=~0$) at low $\mathrm{We}$ is shown in Fig.~\ref{collisions_fig3}a. The neck region (as defined in Fig.~\ref{collisions_fig2}) rapidly grows upon first contact of the droplets due the large curvature near the point of contact. This is accompanied by the formation of capillary waves, which travel over the droplets' interfaces and constructively interfere at the droplets' apexes, forming protrusions. The amplitude and propagation dynamics of the capillary waves on both droplets are equal, since $\gamma_1~=~\gamma_2$ \cite{Keller1983}. The protrusions thus form in phase and with equal amplitudes on both droplets. As a result, the droplet shape remains symmetric during the entire coalescence process. 

Figure~\ref{collisions_fig3}b shows a collision between two identical droplets at high $\mathrm{We}$. Symmetry is maintained, since $\tilde{\gamma}~=~0$. The shape of the droplet, however, is significantly different from the shape of the droplet in Fig.~\ref{collisions_fig3}a---it is more elongated in the vertical direction, and the protrusions have a lower amplitude. The vertical elongation is caused by the higher inertia of the droplets, which, combined with incompressibility of the liquid, forces an outward vertical flow \cite{Ashgriz1990, Roisman2004}. The protrusion amplitude is smaller due to the smaller time scale of the collision (28.5~$\mu$s in  Fig.~\ref{collisions_fig3}a versus 22.0~$\mu$s in Fig.~\ref{collisions_fig3}b). Since capillary waves grow over time, the amplitude of the capillary waves in a high $\mathrm{We}$ collision (which have a smaller time scale) remains smaller than the amplitude of capillary waves in low $\mathrm{We}$ collisions (larger time scale) \cite{Keller1983}. Hence, the protrusion amplitude in Fig.~\ref{collisions_fig3}b is smaller than that in Fig.~\ref{collisions_fig3}a.

We now turn to collisions with $\tilde\gamma~>~0$, as shown in Fig.~\ref{collisions_fig3}c--e. Figure~\ref{collisions_fig3}c shows a collision with high $\tilde{\gamma}$ at low $\mathrm{We}$. A striking difference with respect to the collisions with $\tilde{\gamma}~=~0$ is observed. Namely, the coalescing droplets take on a highly asymmetric shape. A similar asymmetric shape was reported by Gao \emph{et al.} and Kohno \emph{et al.} \cite{Gao2005, Kohno2013, Suzuki2014}. Here, a slender protrusion with large amplitude forms at the side of the water droplet, whereas a small protrusion forms on the ethanol droplet. The difference in protrusion size is caused by the different surface tensions of the droplets. Similarly, the different surface tensions result in different capillary wave velocities, such that the protrusions reach their maximum size out of phase \cite{Keller1983}. The coalescence dynamics and capillary waves are further affected by the Marangoni effect, which causes ethanol to engulf the water droplet \cite{Koldeweij2019}. In some cases, the protrusion pinches-off, such that a satellite droplet is formed, in a process known as `partial coalescence' \cite{Chen2006, Sun2018, Ding2012, Gilet2007, Thoroddsen2000}. However, we typically observe that capillarity acts to pull the protrusion back in, before pinch-off occurs, to restore the droplet to a spherical shape. In section \ref{sec:asymmetricwaves} we study capillary waves in the presence of the Marangoni effect in the limiting case of $\mathrm{We}~=~0$, i.e., in the absence of inertia.  

The asymmetry that forms when $\tilde{\gamma}~>~0$ can be strongly reduced by increasing $\mathrm{We}$, as shown in Fig.~\ref{collisions_fig3}d. The protrusions in Fig.~\ref{collisions_fig3}d are smaller than those in Fig.~\ref{collisions_fig3}c, and are of similar size on both sides of the droplet. We note that, though the droplet shape is not mirror symmetric around the vertical axis, such as in Figs.~\ref{collisions_fig3}a--b, its vertically elongated shape is similar to that of the droplet shown in Fig.~\ref{collisions_fig3}b for $\tilde{\gamma}~=~0$. Key for the strong reduction of the asymmetric droplet shapes (which are caused by asymmetric surface tension) is that the inertia of both droplets is similar, i.e., $\rho_1 {v_{1,x}}^2 R_1~\approx~\rho_2 {v_{2,x}}^2 R_2$. When $\mathrm{We}~\gg~1$, the (symmetric) inertia dominates over the (asymmetric) surface tension, such that the droplet shape remains largely symmetric during coalescence. Finally, Fig.~\ref{collisions_fig3}e shows a collision with moderate $\tilde{\gamma}$ and $\mathrm{We}$, such that the contributions of inertia and surface tension are roughly equal. Indeed, the droplet shape is asymmetric, but less so than in Fig.~\ref{collisions_fig3}c, reinforcing the importance of both $\tilde{\gamma}$ and $\mathrm{We}$ for the droplet shape.

\begin{figure}[t!]
	\begin{center}
		\includegraphics[scale=1.1]{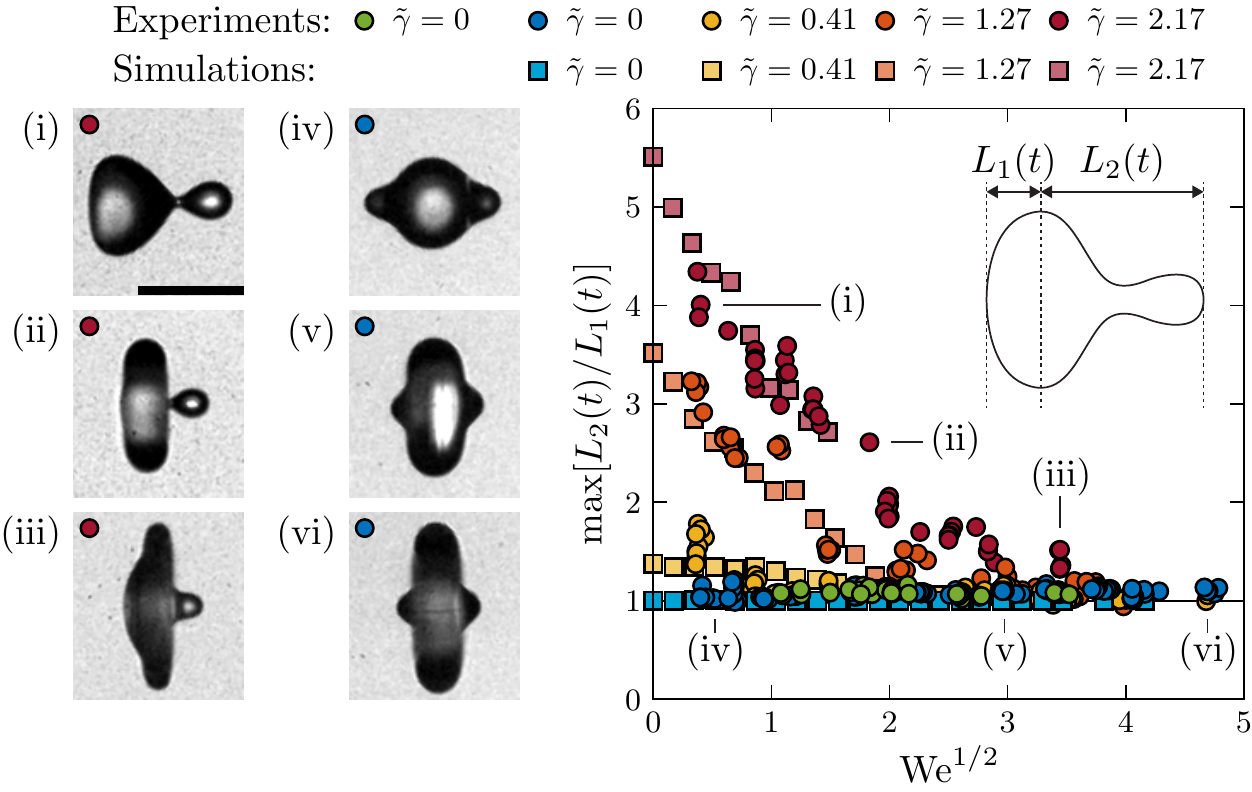}
		\caption{The maximum asymmetry $\mathrm{max}[L_2(t)/L_1(t)]$ as a function of $\mathrm{We}^{1/2}$ for several $\tilde{\gamma}$. The schematic inset shows the definitions of $L_1(t)$ and $L_2(t)$. In these experiments (circles) and simulations (squares) $\gamma_1$ was varied and $\gamma_2~=~72.5$~mN/m was fixed, except for the green symbols, where $\gamma_1~=~\gamma_2~=~22.9$~mN/m. The snapshots (i)--(vi) correspond to the indicated data points. The scalebar corresponds to 100~$\mu$m and applies to all snapshots.}
		\label{collisions_fig4}
	\end{center}
\end{figure}

We further quantify the effects of $\tilde{\gamma}$ and $\mathrm{We}$ on the asymmetry in Fig.~\ref{collisions_fig4}, where we show the maximum asymmetry, which we quantify by the quantity $\mathrm{max}[L_2(t)/L_1(t)]$, as a function of $\mathrm{We^{1/2}}$ for several $\tilde{\gamma}$. Here, $L_1(t)$ and $L_2(t)$ are the time-dependent lengths from the apexes of the two droplets to the point of maximum vertical extension, as defined in the schematic in Fig.~\ref{collisions_fig4}. Their ratio gives a measure for the asymmetry of the droplet shape. Figure~\ref{collisions_fig4} shows the maximum value that the ratio $L_2(t)/L_1(t)$ reaches during the collision. High asymmetry is observed for high $\tilde{\gamma}$, but can be strongly reduced by increasing $\mathrm{We}$ (see snapshots (i)-(iii) in Fig.~\ref{collisions_fig4}). By contrast, droplets with $\tilde{\gamma}~=~0$ always remain symmetric (see snapshots (iv)-(vi)). In line with our observations in Fig.~\ref{collisions_fig3}, we conclude that the asymmetry grows as $\tilde{\gamma}$ increases, and that it decreases with increasing $\mathrm{We}$. Thus, the droplet shape is determined by the competition between the surface tension difference and the inertia of the colliding droplets.

While our results show that asymmetric capillary waves are of paramount importance to the asymmetric droplet shape, the precise influence of the surface tension difference on the capillary waves remains to be determined. For that reason, we systematically study capillary waves in the presence of the Marangoni effect in the next section.

\begin{figure}[b!]
	\begin{center}
		\includegraphics[scale=1.1]{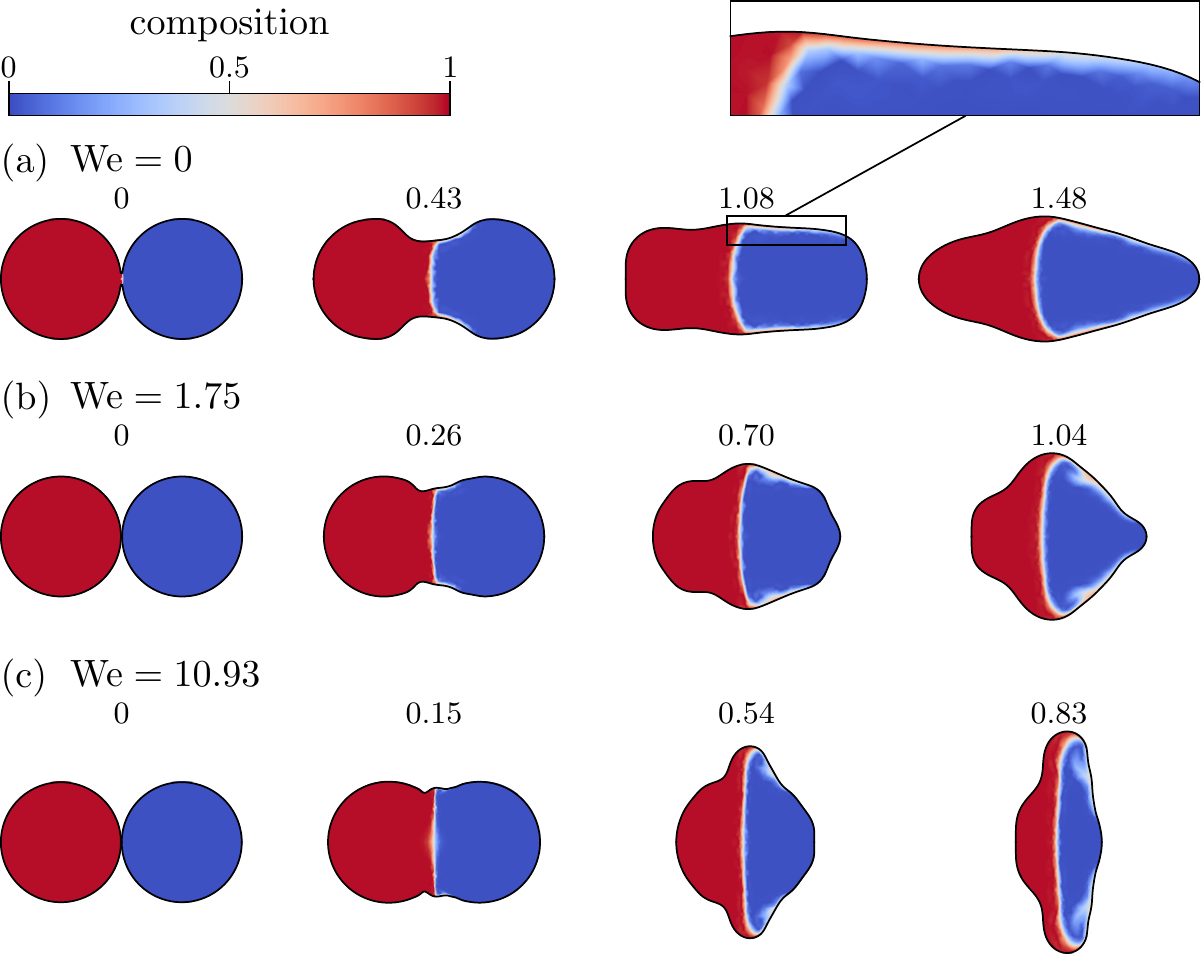}
		\caption{Snapshots from the numerical simulations with $\tilde{\gamma}~=~1$ and various $\mathrm{We}$. Here, $\rho~=~1000$~kg/m$^3$, $\eta~=~1$~mPa$\cdot$s, $\gamma_1~=~40$~mN/m, and $\gamma_2~=~80$~mN/m. The numbers above the snapshots indicate the non-dimensional time $\tilde{t}~=~t/\tau_2$, with $\tilde{t}~=~0$ being the moment of first contact between the droplets. The zoom provides a closer look at the engulfing film. The protrusion amplitude (and the collision time scale) are smaller for higher $\mathrm{We}$. See the Supplementary Material for the corresponding movies.}
		\label{collisions_fig5}
	\end{center}
\end{figure}

\begin{figure}[b!]
	\begin{center}
		\includegraphics[scale=1.1]{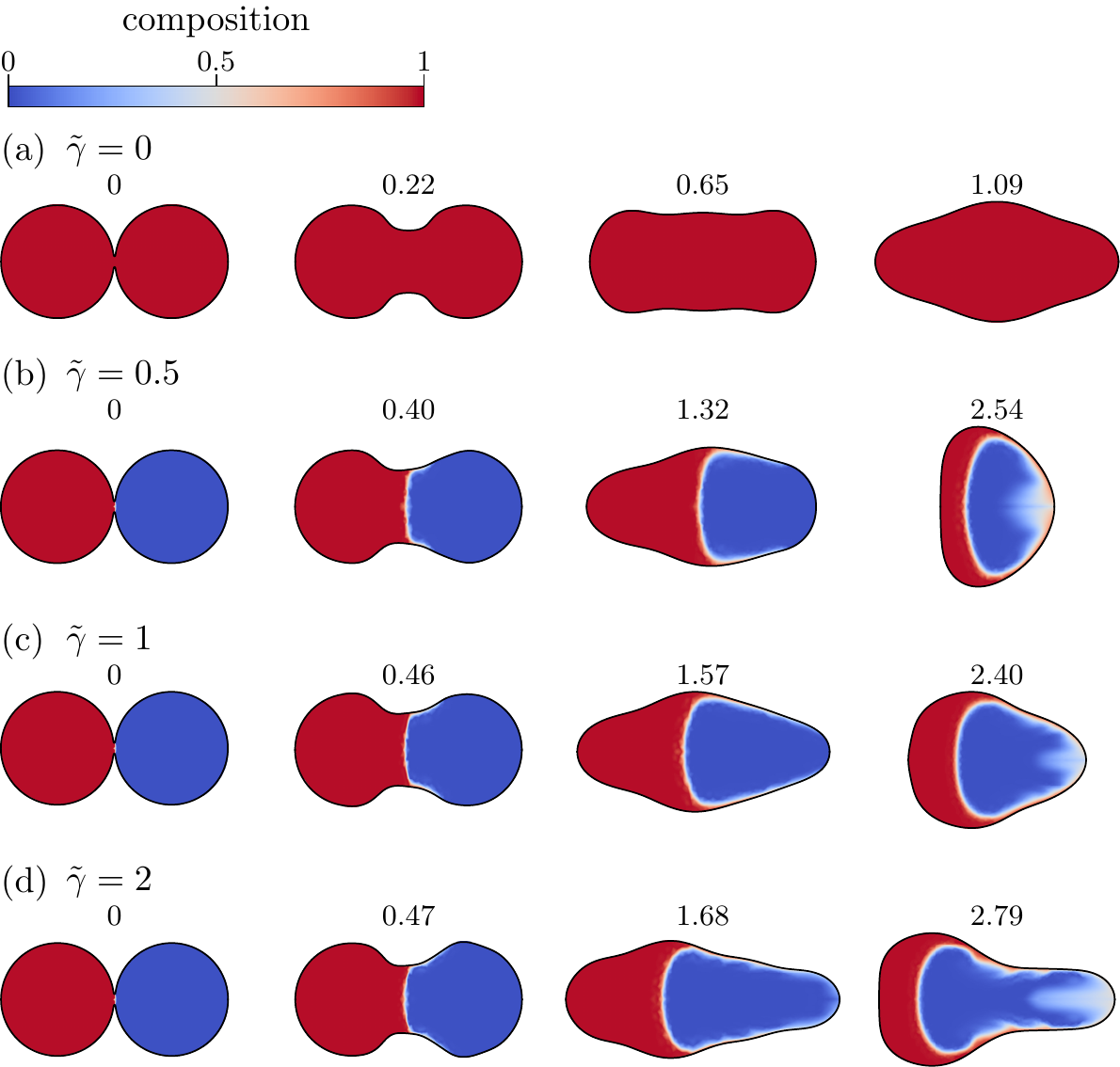}
		\caption{Snapshots from the numerical simulations with $\mathrm{We}~=~0$ and various $\tilde{\gamma}$. Here, $\rho~=~1000$~kg/m$^3$, $\eta~=~1$~mPa$\cdot$s, $\gamma_1~=~20$~mN/m, and $\gamma_2$ is varied between 20~mN/m and 60~mN/m, resulting in the shown values of $\tilde{\gamma}$. The numbers above the snapshots indicate $\tilde{t}~=~t/\tau_2$. The protrusion amplitude increases with increasing $\tilde{\gamma}$, resulting in highly asymmetric droplet shapes. See the Supplementary Material for the corresponding movies.}
		\label{collisions_fig6}
	\end{center}
\end{figure}

\section{Asymmetric capillary waves}
\label{sec:asymmetricwaves}
Asymmetric capillary waves form when two droplets of different surface tensions coalesce. In the previous section, we showed that this leads to asymmetric droplet shapes. Here, we focus solely on the dynamics of the  capillary waves, which we study using numerical simulations of a model system. It will turn out that the capillary waves, and not the Marangoni effect, are responsible for the observed droplet asymmetry. 

In the model system we impose that all properties of the droplets are equal, except their surface tensions. Specifically, we set $\rho~=~1000$~kg/m$^3$, $\eta~=~1$~mPa$\cdot$s, $D~=~1\times10^{-9}$~m$^2$/s, $R~=~35$~$\mu$m, and use a linear model for the surface tension $\gamma = \gamma_2+(\gamma_1-\gamma_2)c$, where $c$ varies between 0 and 1, and $\gamma_1~\geq~20$~mN/m and $\gamma_2~\leq~100$~mN/m. While this constitutes a simplified model, the values used here are close to those of a water-ethanol system, and the dimensionless groups are of the same order of magnitude for this model and the experiments presented before.   

\begin{figure}[t!]
	\begin{center}
		\includegraphics[scale=1.1]{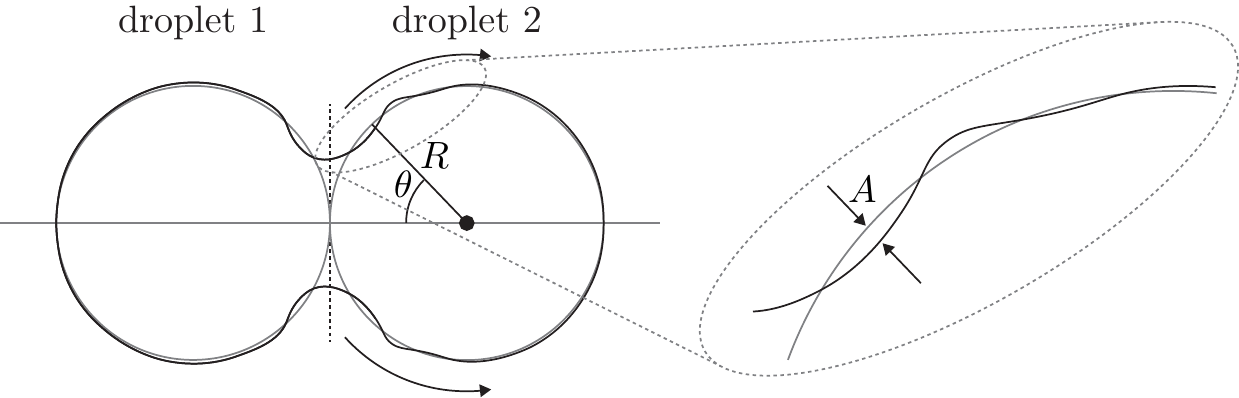}
		\caption{Schematic of the capillary waves on the droplets. The amplitude $A$ of the capillary waves is determined with respect to the shape of the droplet at $t~=~0$ as a function of $\theta$, which is the angle between an arbitrary point on the droplet interface and the center of the droplet at $t~=~0$. In this case we neglect the small initial connection between the droplets, and assume that the droplets are perfect circles. The zoom shows the definition of $A$.}
		\label{collisions_fig7}
	\end{center}
\end{figure}

Figure~\ref{collisions_fig5} shows example snapshots of these simulations for various $\mathrm{We}$. A thin film of the low surface tension liquid is observed to engulf the high surface tension droplet during the coalescence process, due to the Marangoni effect \cite{Koldeweij2019}. This induces tangential stresses at the interface which are expected to change the capillary wave dynamics and shape. For example, when a low surface tension droplet is deposited on a liquid substrate of higher surface tension, these stresses can induce a local interface distortion know as the `Marangoni ridge' \cite{HernandezSanchez2015, Wang2015}.

In the remainder of the paper, we further simplify the problem by excluding inertia, i.e., we set $\mathrm{We}~=~0$. Example snapshots of such simulations are shown in Fig.~\ref{collisions_fig6}. The simulations reveal that the main morphological change of the droplet shape takes place on the droplet with high surface tension, where the capillary waves take on a different shape, depending on the value of $\tilde{\gamma}$. By contrast, the low surface tension droplet (with $\gamma_1~=~20$~mN/m fixed for the simulations shown in Fig.~\ref{collisions_fig6}) appears to be relatively unaffected by changes in $\tilde{\gamma}$. Additionally, we note that, despite their miscibility, the two liquids do not strongly mix during the coalescence process, in line with previous observations for a similar system \cite{Suzuki2014}.

\begin{figure}[t!]
	\begin{center}
		\includegraphics[scale=1.1]{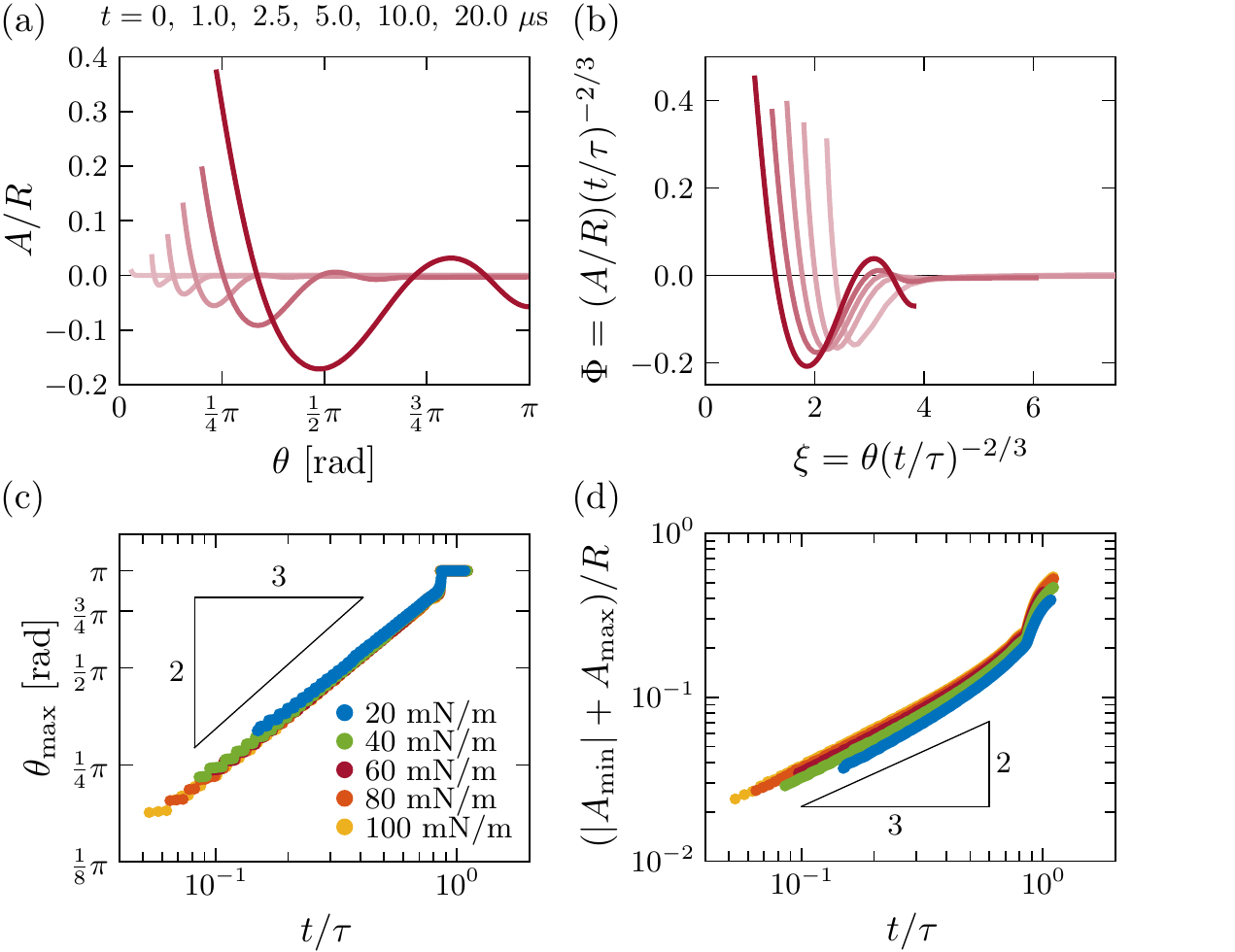}
		\caption{Capillary waves on coalescing droplets for $\tilde{\gamma}~=~0$ and $\mathrm{We}~=~0$, from our numerical simulations. (a) Capillary waves on a droplet with $\gamma~=~60$~mN/m, $A$ is the amplitude and $\theta$ the angular coordinate as defined in Fig.~\ref{collisions_fig7}. (b) Similarity collapse of the capillary waves, following Ref.~\cite{Keller1983}. (c) The location of the capillary wave maximum as a function of time. (d) The amplitude of the capillary wave as a function of time.}
		\label{collisions_fig8}
	\end{center}
\end{figure}

To better understand the influence of $\tilde{\gamma}$ on the dynamics of the capillary waves, we study the shape of the capillary waves over time with respect to the original droplet shape at $t~=~0$. Figure~\ref{collisions_fig7} shows the method used to extract the capillary wave amplitude. A similar method was successfully used by Thoroddsen \emph{et al.} for millimetere-sized drop coalescence \cite{Thoroddsen2007}. Figure~\ref{collisions_fig8}a shows the capillary waves that appear on coalescing droplets in the absence of a surface tension difference ($\tilde{\gamma}~=~0$, $\mathrm{We}~=~0$). In Fig.~\ref{collisions_fig8}b, we attempt to collapse the capillary waves using the similarity scaling proposed by Keller \& Miksis for capillary waves on flat surfaces \cite{Keller1983}. We do not find perfect collapse of the data, probably due to the high curvature of the droplet interface. However, in Fig.~\ref{collisions_fig8}c--d we report the location of the capillary wave maximum and the amplitude of the capillary waves as a function of time, finding that their dynamics are indeed close to the expected $t^{2/3}$ scaling \cite{Keller1983}.

\begin{figure}[t!]
	\begin{center}
		\includegraphics[scale=1.1]{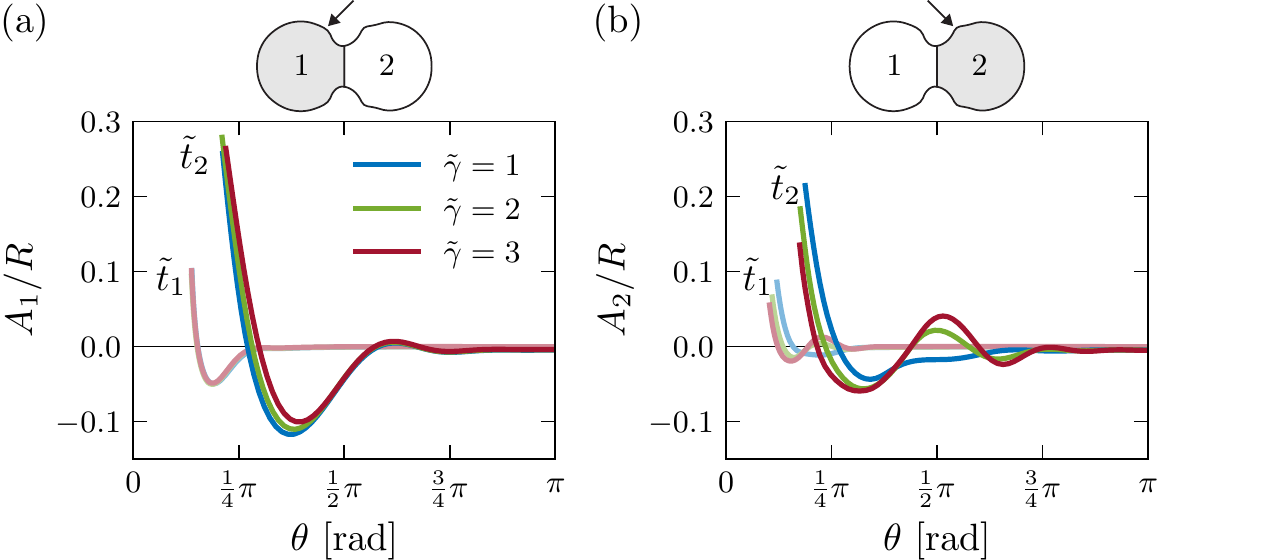}
		\caption{Capillary waves for various $\tilde{\gamma}$ at $\tilde{t}_1~=~0.15$ and $\tilde{t}_2~=~0.50$, from our numerical simulations. Here, $\gamma_1~=~20$~mN/m, and $\gamma_2~=~40,~60,~80$~mN/m, resulting in $\tilde{\gamma}~=~1,~2,~3$. (a) Capillary wave amplitude on the low surface tension droplet ($\tilde{t}~=~t/\tau_1$). (b) Capillary wave amplitude on the high surface tension droplet ($\tilde{t}~=~t/\tau_2$).}
		\label{collisions_fig9}
	\end{center}
	\begin{center}
		\includegraphics[scale=1.1]{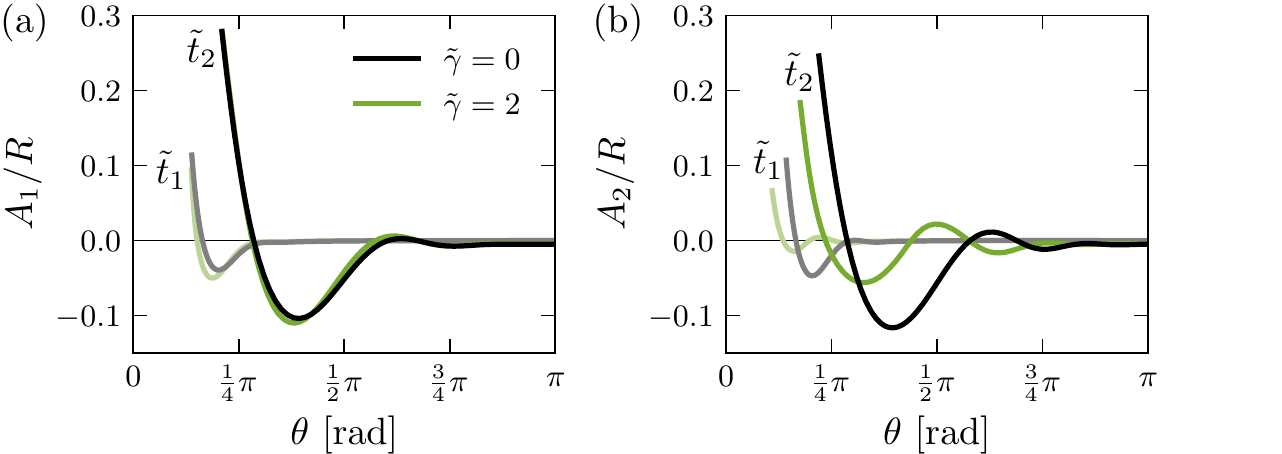}
		\caption{Capillary waves for various $\tilde{\gamma}~=~0$ and $\tilde{\gamma}~=~2$ at $\tilde{t}_1~=~0.15$ and $\tilde{t}_2~=~0.50$, from our numerical simulations. (a) Capillary wave amplitude on the low surface tension droplet ($\tilde{t}~=~t/\tau_1$), with $\gamma_1~=~20$~mN/m. The black line indicates capillary waves on a droplet with $\gamma~=~20$~mN/m. (b) Capillary wave amplitude on the high surface tension droplet ($\tilde{t}~=~t/\tau_2$), with $\gamma_2~=~40$~mN/m. The black line indicates capillary waves on a droplet with $\gamma~=~40$~mN/m.}
		\label{collisions_fig10}
	\end{center}
\end{figure}

We now return to coalescence with $\tilde{\gamma}~>~0$. The capillary waves that form on the droplets for several values of $\tilde{\gamma}$ are shown in Fig.~\ref{collisions_fig9}. The amplitude of the capillary waves on the low surface tension droplet (droplet 1) is shown in Fig.~\ref{collisions_fig9}a for two different non-dimensional times $\tilde{t}_1$ and $\tilde{t}_2$, where $\tilde{t}~=~t/\tau_1$ with $\tau_1~=~(\rho R^3/\gamma_1)^{1/2}$. The surface tension $\gamma_1~=~20$~mN/m for the three cases shown in Fig.~\ref{collisions_fig9}a, and remains constant during the coalescence process, since the Marangoni flow is directed from droplet 1 to droplet 2. We observe that $\tilde{\gamma}$ has almost no effect on the capillary waves on droplet 1---their amplitude and propagation dynamics are nearly the same for all $\tilde{\gamma}$. 

\begin{figure}[b!]
	\begin{center}
		\includegraphics[scale=1.1]{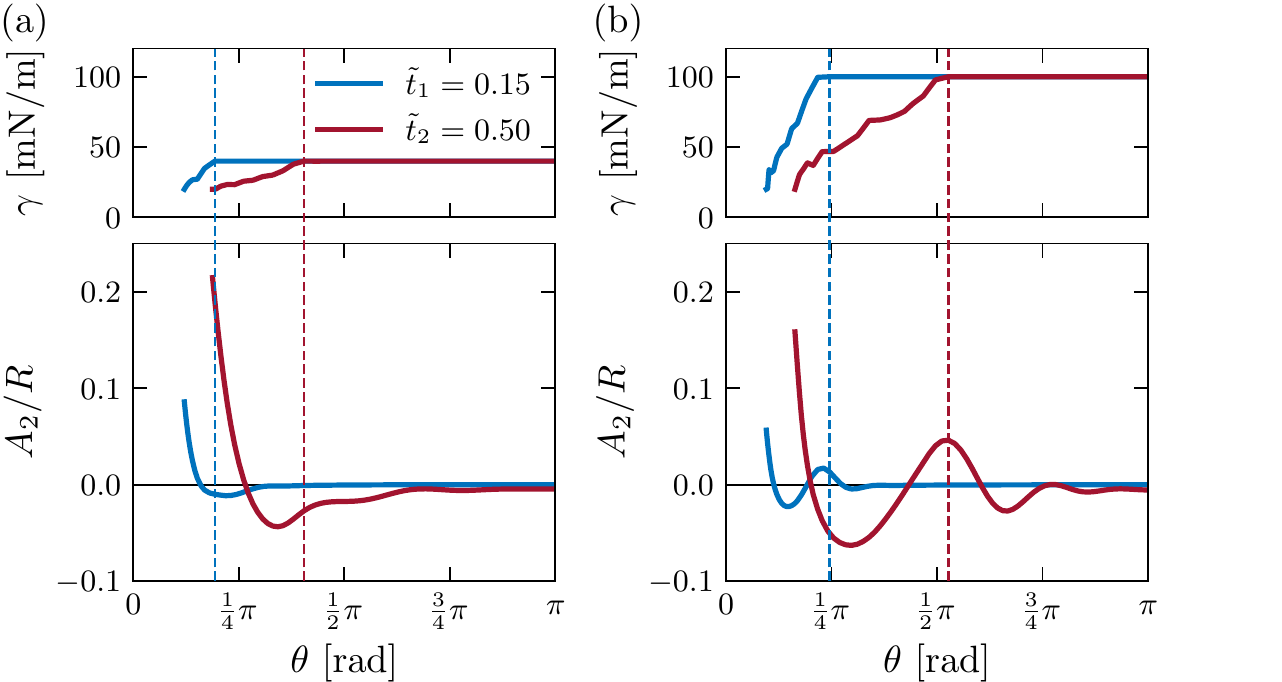}
		\caption{Surface tension and relative capillary wave amplitude on droplet 2 from our numerical simulations for (a) $\tilde{\gamma}~=~1$, and (b) $\tilde{\gamma}~=~4$. In both cases $\gamma_1~=~20$~mN/m. The vertical dashed lines indicate the engulfing front locations.}
		\label{collisions_fig11}
	\end{center}
\end{figure}

Figure~\ref{collisions_fig9}b shows the corresponding capillary waves on droplet 2 at $\tilde{t}_1$ and $\tilde{t}_2$, where the time is now normalized by $\tau_2~=~(\rho R^3/\gamma_2)^{1/2}$. Several significant effects  of $\tilde{\gamma}$ on the capillary waves on droplet 2 can be observed. First, the capillary waves are strongly deformed by the presence of the Marangoni effect. Compare, for example, the shape of the first local maximum (at $\theta~\approx~\pi/2$ for $\tilde{t}_2$), which appears damped and distorted for $\tilde{\gamma}~=~1$. One would expect that the amplitude of the capillary waves on droplet 2 ($A_2$) to be larger than the amplitude of the capillary waves on droplet 1 ($A_1$), since $\gamma_2~>~\gamma_1$, yet, when we compare $A_1$ (Fig.~\ref{collisions_fig9}a) to $A_2$ (Fig.~\ref{collisions_fig9}b), we find the inverse, at least for the amplitude of the first local minimum. Second, the propagation of the capillary waves is slower for increasing $\tilde{\gamma}$. The higher $\tilde{\gamma}$ waves are consistently lagging behind the $\tilde\gamma~=~1$ wave in Fig.~\ref{collisions_fig9}b. It is, however, important to note that $\gamma_2$ is not identical for the three cases considered in Fig.~\ref{collisions_fig9}b. The capillary waves in Fig.~\ref{collisions_fig9}b are therefore not expected to collapse, since their properties depend on $\gamma_2$ \cite{Keller1983}. We therefore compare the capillary waves for $\tilde{\gamma}~=~2$ to those for $\tilde{\gamma}~=~0$ in Fig.~\ref{collisions_fig10}, to verify that the difference between the capillary waves in Fig.~\ref{collisions_fig9}b is (at least partly) caused by $\tilde{\gamma}~\neq~0$ and not solely by the difference in $\gamma_2$.  In Fig.~\ref{collisions_fig10}a we compare the wave amplitude $A_1$ for the two cases with $\tilde{\gamma}~=~0$ and $\tilde{\gamma}~=~2$, where for both cases $\gamma_1~=~20$~mN/m. In line with the results shown in Fig.~\ref{collisions_fig9}a, there is no effect of $\tilde{\gamma}$ on the wave amplitude $A_1$. In Fig.~\ref{collisions_fig10}b we show a similar comparison for the capillary waves on droplet 2, where we note that $\gamma_2~=~40$~mN/m for both cases. Here, we once again observe that the damping increases and wave propagation decreases with increasing $\tilde{\gamma}$. We note that a similar damping effect on capillary waves has recently been observed for the coalescence of a surfactant-laden droplet coalescing with a liquid bath \cite{Constante-Amores2021}. We thus conclude that $\tilde{\gamma}$ has a strong effect on the capillary waves (and by extent, the protrusion) on droplet 2. Additionally, the damping effect of $\tilde{\gamma}$ indicates that the Marangoni effect does not drive the asymmetric droplet shapes observed in Figs.~\ref{collisions_fig3} and \ref{collisions_fig4}, in contrast to the findings of Gao \emph{et al.} and Kohno \emph{et al.} \cite{Gao2005, Kohno2013, Suzuki2014}. In fact, the intrinsic dependence of the capillary wave dynamics on surface tension drives the asymmetry, and remarkably, the Marangoni effect decreases the asymmetry. 

\begin{figure}[t!]
	\begin{center}
		\includegraphics[scale=1.1]{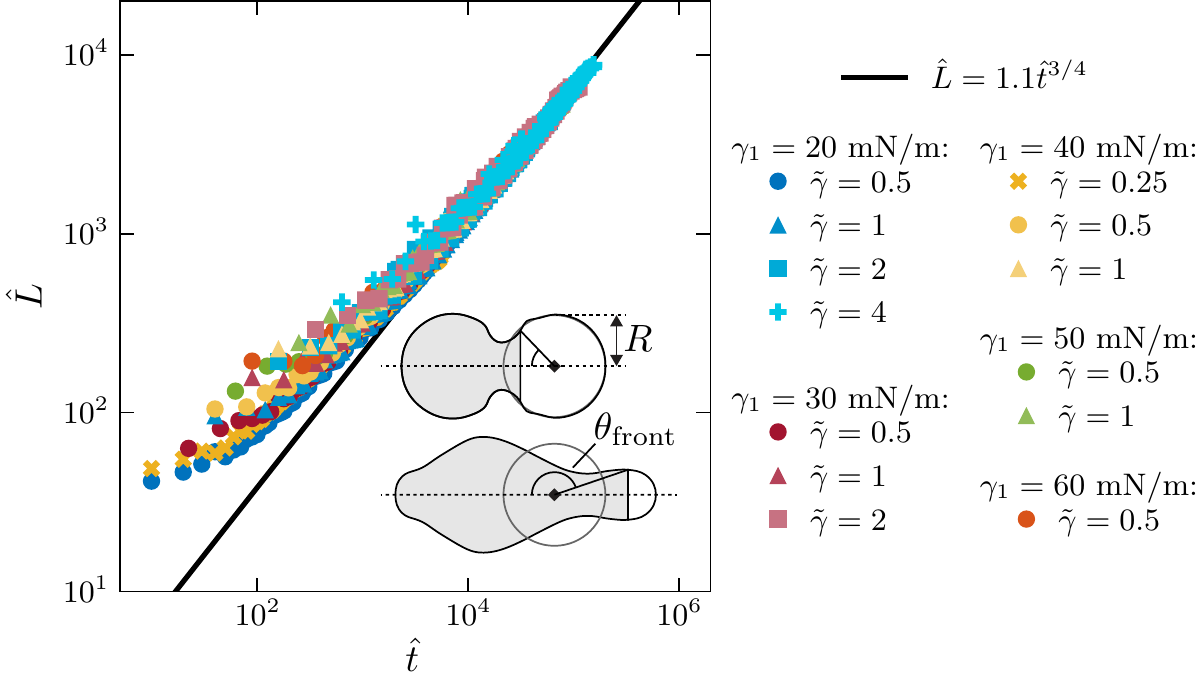}
		\caption{Rescaled location of the engulfing front as a function of rescaled time, from our numerical simulations. The schematic inset shows the definition of $\theta_\mathrm{front}$, which is the angle between the location of the engulfing front on the droplet interface with respect to the center of the initial droplet position at $t~=~0$. The data are cut off when $\theta_\mathrm{front}~=~\pi$. The front dynamics converge to  $\hat{L} \propto t^{3/4}$, in accordance with the results of Koldeweij \emph{et al.} \cite{Koldeweij2019}.}
		\label{collisions_fig12}
	\end{center}
\end{figure}

Figures~\ref{collisions_fig9}b and \ref{collisions_fig10}b show that the capillary waves on the high surface tension droplet are strongly affected by the engulfing front of the low surface tension liquid. However, it is not yet clear if the distortion of the capillary wave is universal. In Fig.~\ref{collisions_fig11} we therefore compare the local surface tension to the capillary wave amplitude as a function of $\theta$ for two $\tilde{\gamma}$. The dashed line in Fig.~\ref{collisions_fig11} indicates the location of the engulfing front, which we define as the first $\theta$ where $\gamma~<~\gamma_2$. Figure~\ref{collisions_fig11} reveals that the location of the engulfing front with respect to the capillary wave is not universal, since the location of the front with respect to the capillary wave is different in Fig.~\ref{collisions_fig11}a and b. This implies that the dynamics of the capillary waves (characterized by $\theta_\mathrm{max} \propto t^{2/3}$) are different than that of the engulfing front. To further quantify this point, we therefore track the front location with respect to the initial droplet position at $t~=~0$ in Fig.~\ref{collisions_fig12}. Following Koldeweij \emph{et al.} \cite{Koldeweij2019, Huh1975}, we rescale the time and front location using
\begin{equation}
\hat{t} = \frac{t}{\eta^3/\Delta\gamma^2\rho}, \quad \quad \hat{L} = \frac{\theta_\mathrm{front} R}{\eta^2/\Delta\gamma\rho}, \quad \quad 
\end{equation}
where $\Delta\gamma~=~\gamma_2-\gamma_1$. While the data collapses at large $\hat{t}$, deviations can be observed for small $\hat{t}$ due to the initial conditions of our simulations. We find that the dynamics converge to $\hat{L} \propto t^{3/4}$. The same exponent was found for ethanol spreading over millimeter-sized drops in the pendant geometry \cite{Koldeweij2019}. It follows from balancing the Marangoni forces and the viscous forces in the thin boundary layer of the spreading liquid \cite{Hoult1972, Foda1980, Berg2009}. We thus find that the engulfing front exhibits different dynamics ($\propto~t^{3/4}$) than the capillary wave ($\propto~t^{2/3}$), reflecting the different driving mechanisms.This explains why the shape of capillary waves in the presence of the Marangoni effect is non-universal. 

\section{Conclusion}
\label{sec:conclusion}
In this work, we have studied collisions between two droplets with different surface tensions. Using a combination of experiments and numerical simulations, we have shown that the shape of the droplets during coalescence can be highly asymmetric. By contrast, the shape of colliding droplets with identical surface tensions remains symmetric at all times. Furthermore, we have shown that the droplet shape is determined by a delicate competition between the surface tension difference and inertia---the asymmetry increases with increasing surface tension difference, and can be reduced by increasing the inertia of the colliding droplets. 

Our results show that the interference of capillary waves is at the origin of the asymmetric droplet shapes. Since the amplitude and propagation dynamics of capillary waves depend on the surface tension of the interface that they travel on, the capillary waves are asymmetric in the coalescence of two droplets with different surface tensions. Contrary to previous works, we find that the Marangoni effect has a damping effect on the capillary waves \cite{Gao2005, Kohno2013, Suzuki2014}. Hence, we find that the asymmetry is primarily caused by the intrinsic difference between the capillary waves amplitude and propagation dynamics set by the initial surface tension of the droplets, and not by the Marangoni effect. In fact, the Marangoni effect decreases the asymmetry, by damping the capillary waves.

The results presented here show the richness of phenomena which can occur in physicochemical hydrodynamics \cite{Lohse2020}, and that these are often counterintuitive. From a more applied point of view, they may be of interest to microfluidic applications where coalescence of droplets with different surface tensions is prevalent, such as the in-air microfluidic fabrication of emulsions \cite{Visser2018},  colliding-droplet chemical microreactors \cite{Davis2017}, or in other applications of physicochemical hydrodynamics. 

\section*{Acknowledgements}
The authors thank Claas Willem Visser for fruitful discussions. This work is part of an Industrial Partnership Programme (IPP) of the Netherlands Organization for Scientific Research (NWO). This research programme is cofinanced by Canon Production Printing Netherlands B.V., University of Twente and Eindhoven University of Technology. D.L. acknowledges support from the ERC Advanced Grant “DDD” under the project number 740479.

\nocite{*}

\bibliography{manuscriptBib}

\end{document}